\documentclass[aps,prl,epsfigure,showpacs,twocolumn]{revtex4}
\usepackage{graphicx}
\usepackage{amsmath}
\usepackage{amstext}
\usepackage{latexsym}
\usepackage{amsfonts}
\usepackage{bm}
\usepackage{amssymb}
\usepackage{amscd}
\begin{document}
\title{Reply to Q. Zhang et al.}
\author{Won-Ho Kye}
\affiliation{The Korean Intellectual Property Office, Daejeon
302-701, Korea}
\author{M. S. Kim}
\affiliation{School of Mathematics and Physics, Queen's
University, Belfast BT7 1NN, United Kingdom}
 \maketitle

Recently, there have been a considerable interest in a quantum
key distribution (QKD) developed by us, which we call the KKKP protocol
due to initials of the authors\cite{Kye}. The KKKP protocol is based on
random polarizations and three-way communications between Alice
and Bob, two legitimate users of the key.
In \cite{Kye} we extended the KKKP protocol in order to make it robust
against the impersonation attack by employing a set of two pulses
to embody a qubit.
Bob puts
his private information $s$ by random shuffling and Alice puts
her private information $b$ by random blocking.  Then Bob's final
measurement outcome $l$ depends on Alice's key $k$ and those
private information $s$ and $b$: $l=s\oplus b\oplus\ k$.
Impersonating Eve can get $b\oplus k$ but she does not know the
shuffling parameter $s$ so that there should be an error in Bob's
measurement outcome.
However, Zhang {\it
et al.} recognized that because the shuffling factors of the
first and second pulses are strongly correlated (the former being
$s$ and the latter being $s\oplus1$), impersonating Eve does not
need to know $s$ to send correct information to Bob after reading
$b\oplus k$.  Then, when the blocking factor $b$ is announced
through the public channel, Eve gets the keys without causing any
error to Bob's reading of the key.

Here, we slightly modify
the KKKP protocol for Bob to give two independent shuffling factors to the
first and second pulses respectively. In this way, we do not lose
all the advantages and basic philosophy of the KKKP protocol while we
build its security against the impersonation attack:

\noindent
{\bf (Q.1)} Alice prepares two qubits in
    $
        |\psi_1\rangle = |\theta_1\rangle \otimes |\theta_2\rangle.
    $
\noindent
{\bf (Q.2)} Upon reception of the two qubits, Bob applies random shuffling
$\hat{U}_y(\phi +(-1)^{s_1}\pi/4) \otimes \hat{U}_y(\phi
+(-1)^{s_2}\pi/4)$, where $s_i=\{0, 1\}~,~(i=1,2)$, are two
independent random numbers.  He sends the qubits back to Alice.

\noindent
{\bf (Q.3)} Upon reception of the pulses, Alice applies $\hat{U}_y(-\theta_1
+ (-1)^{k} \pi/4)\otimes\hat{U}_y(-\theta_2 + (-1)^{k\oplus
1}\pi/4)$ where $k\in \{0, 1\}$ is the key bit. Alice blocks one
of the qubits, after which the surviving qubit is given by
\begin{equation}
|\phi+(-1)^{s_b}\pi/4+(-1)^{k\oplus b \oplus 1}\pi/4\rangle,
\end{equation}
where $b$ is the blocking factor to let the first ($b=1$) or the
second pulse ($b=2$) go.

\noindent
{\bf (Q.4)} Bob receives the qubit and applies  $\hat{U}_y(-\phi)$ on them
before he measures it. The measurement outcome is given by $l=s_b
\oplus k \oplus b$. The key  is given by $k = s_b\oplus l \oplus
b$.

\noindent
\%item [(Q.5)] 
{\bf (Q.5)} After repeating N times from (Q.1) to (Q.4),
Alice announces blocking factors $b$ through a public channel and Bob verifies the
shared key by exchanging the hash value of the key.

Now, we show that the attack proposed by Zhang et al. is easily noticed in the
QKD modified as above.

\noindent
{\bf (Q.1$^\prime$)} After (Q.1), Eve intercepts and stores  the both pulses from
Alice in ``set E1''. Thus Eve has E1=$\{|\theta_1\rangle \otimes
|\theta_2\rangle\}$. Eve sends to Bob two pulses originally
prepared by her with random angles $\theta_1^\prime,~
\theta_2^\prime$.

\noindent
{\bf (Q.2$^\prime$)} After step (Q.2), Eve intercepts both pulses from Bob and stores
them in ``set E2'' after compensating with the angles
$-\theta_1^\prime,~-\theta_2^\prime$. Eve then has
E2=$\{|\phi+(-1)^{s_1}\pi/4\rangle \otimes |\phi + (-1)^{s_2}
\pi/4\rangle \}$. Eve needs to guess two random parameters $s_1$
and $s_2$. Consider that Eve chooses her shuffling parameters
$s_1^\prime=s_2 ^\prime=0$ (this is one possibility out of
four.). Eve shuffles E1 which becomes
E1$^{\prime}=\{|\theta_1+(-1)^{s_1^\prime}\pi/4\rangle \otimes
|\theta_2 +(-1)^{s_2^\prime}\pi/4\rangle\}$, and sends it to
Alice.

\noindent
{\bf (Q.3$^\prime$)} After step (Q.3), Eve intercepts the returning qubit
$|(-1)^{s_b^\prime}\pi/4 + (-1)^{k\oplus b \oplus 1}\pi/4\rangle$
and measures it to read the pre-key value
$l^\prime=s_b^\prime\oplus k \oplus b= k\oplus b$ because
$s_1^\prime=s_2^\prime=0$. She then encodes $(-1)^{k\oplus
b\oplus 1}\pi/4$  onto one of E2. If Eve takes the first qubit,
the qubit becomes $|\phi+(-1)^{s_1}\pi/4 + (-1)^{k\oplus b\oplus
1}\pi/4\rangle$ and Bob measures $l^\prime_1=s_1\oplus k \oplus
b$.  Otherwise, Bob measures $l^\prime_2=s_2\oplus k \oplus b$.
Regardless $l^\prime_1$ or $l^\prime_2$, there would be a  25\%
error rate with $l=s_b\oplus k \oplus b$.  This should be easily
noticed in (Q.5).


We have proved that the slightly modified KKKP protocol becomes robust
against the impersonation attack. One important point is that
Alice should give special care not to give a chance for Eve to
find the blocking factor before Eve returns encoded qubits to Bob
in the step (Q.3$^\prime$). Here, Eve may try to use spy pulses of
different frequencies or different intensities in order to find
this information. This kind of attempt should be filtered out by
a careful design of the setup\cite{Gisin}.  For the case of coherent state implementation,
Alice can randomly check if the two pulses are of the same amplitude by sending them to a 50:50 beam splitter. 
When they are identical, all the photons should be detected at only one output port. If Eve uses two different pulses to get $b$, her action will be detected by Alice.

 \acknowledgments {\it Acknowledgments}- We thank Prof. W. Y.
Hwang for discussions.

\end{document}